\documentclass[]{aa}
\pdfoutput=1
\usepackage{graphicx}
\usepackage{amsmath,amsfonts,amssymb,tabu}
\usepackage[para]{threeparttable}
\usepackage{txfonts}
\usepackage[breaklinks,colorlinks,citecolor=blue,pdfa=true,backref=page]{hyperref}
\usepackage{color}
\usepackage{fixltx2e}
\usepackage{natbib,twoopt}
\usepackage{url}
\usepackage{multirow}
\usepackage{epsf}
\usepackage{epsfig}
\usepackage{longtable}
\usepackage{float}
\usepackage{subfig}
\usepackage{caption}
\newcommand{\GG}[1]{}

\definecolor{mypink1}{RGB}{219, 48, 122}

\usepackage{ifthen}
\usepackage[T1]{fontenc}
\usepackage{lmodern}
\usepackage{ifxetex,ifluatex}
\usepackage{latexsym}
\usepackage{pdfpages}
\usepackage{dblfloatfix}
\usepackage{morefloats}
\usepackage{caption}
\usepackage{lscape}
\usepackage{mathtools}

\bibpunct{(}{)}{;}{a}{}{,} 
\makeatletter
\newcommandtwoopt{\citeads}[3][][]{\href{http://adsabs.harvard.edu/abs/#3}%
{\def\hyper@linkstart##1##2{}%
\let\hyper@linkend\@empty\citealp[#1][#2]{#3}}}
\newcommandtwoopt{\citepads}[3][][]{\href{http://adsabs.harvard.edu/abs/#3}%
{\def\hyper@linkstart##1##2{}%
\let\hyper@linkend\@empty\citep[#1][#2]{#3}}}
\newcommandtwoopt{\citetads}[3][][]{\href{http://adsabs.harvard.edu/abs/#3}%
{\def\hyper@linkstart##1##2{}%
\let\hyper@linkend\@empty\citet[#1][#2]{#3}}}
\newcommandtwoopt{\citeyearads}[3][][]%
{\href{http://adsabs.harvard.edu/abs/#3}
{\def\hyper@linkstart##1##2{}%
\let\hyper@linkend\@empty\citeyear[#1][#2]{#3}}}
\makeatother

\usepackage{graphicx}

\def\simi   {$\sim$\,}

\def \deg         {\text{$^{\circ}$}}
\def \arcmin      {\text{$^\prime$}}
\def \arcsec      {\text{$^{\prime\prime}$}}

\begin{document}

   \title{X-shaped radio galaxy 3C 223.1: A `double boomerang' with an anomalous spectral gradient}

   \titlerunning{XRG 3C223.1}

\author {Gopal-Krishna\inst{1}\thanks{E-mail: gopaltani@gmail.com}
\and Pratik Dabhade\inst{2}\thanks{E-mail: pratik.dabhade@obspm.fr}
}
  
\institute{$^{1}$UM-DAE Centre for Excellence in Basic Sciences (CEBS), Vidyanagari, Mumbai - 400098, India\\ 
$^{2}$Observatoire de Paris, LERMA, Coll\`ege de France, CNRS, PSL University, Sorbonne University, 75014, Paris, France\\
}
 \vspace{-0.5cm}

 \date{\today} 
 \abstract{
 A comparison of the recent LOFAR 144 MHz map of the radio source 3C 223.1 (J094124.028+394441.95) with the VLA maps at 4.9 GHz and 8.3 GHz that we built based on archival data, establishes this X-shaped radio galaxy (XRG) as a singularly robust case where the `wings' exhibit a distinctly flatter radio spectrum than the primary lobes. The details of its anomalous spectral gradient are unravelled here with unprecedented precision. We also highlight the `double boomerang' type radio morphology of this XRG.  It appears plausible that the peculiar spectral gradient in this XRG is owed to  particle acceleration associated with the rebounding of the collimated backflows of synchrotron plasma streaming through its two primary lobes, as they impinge upon and encounter the magnetic tension in the prominent dusty disk of the elliptical galaxy hosting this XRG. We also draw attention to an intriguing new morphological peculiarity among XRGs, namely, a lateral offset observed between the (parallel) axes of the two primary radio lobes.}
 


\maketitle

\section{Introduction} \label{sec:intro}
X-shaped radio galaxies (XRG) are a numerically small but enigmatic species in the zoo of radio galaxies, since their radio emission arises from not one, but two (mis-aligned) pairs of radio lobes of a comparable extent \citep[e.g.][]{leahy84,Capetti02}. Of these, the `primary' (i.e. active) lobes often show a terminal hot spot that signifies an ongoing energy supply via bi-polar jets. In contrast, the secondary lobes (`wings') are usually diffuse and devoid of a terminal hot spot. Two major explanations have been advanced for this morphological dichotomy: (i) each wing is merely a continuation of the hydrodynamical `back flow' in the primary lobe, which gets deflected due to buoyancy forces, upon impinging on an ellipsoidal hot interstellar medium (ISM) of the parent galaxy \citep{leahy84,Worrall95,Hodges-Kluck10}; or (ii) the wings are relics of the lobe pair whose energy supply ceased as the twin-jets feeding them flipped over in a new direction due to a merger of the jetted super-massive black hole (SMBH) with another SMBH, thus giving rise to the active lobes seen presently  \citep{Rottmann01,Zier01}. This possibility of a spin-flip via the SMBH merger and consequent emission of gravity waves \citep{Merritt02} brought XRGs into the limelight about two decades ago, even though the first example of an XRG (3C 315) has been known for nearly half a century \citep{hogbom74}. 

Clearly, spectral index mapping as an indicator of the ages of different parts of XRGs is a key step towards understanding the origin of the XRG phenomenon. Early studies of XRG 3C 315 resulted in contradictory claims about spectral index gradients in this radio source \citep{Hogbom79,Alexander87}. The reported lack of spectral gradients \citep{Hogbom79} was intriguing, since in both above models of XRGs, the wings are identified as the repository of aged synchrotron plasma. \citet{Rottmann01} investigated this issue by comparing his single-dish (Effelsberg) images of nine prominent XRGs at 10.5 GHz (beam \simi 1.15\arcmin) and seven of them also at 32 GHz (beam \simi0.45\arcmin) with the existing Westerbork telescope maps made below 1 GHz as well as VLA maps between 1 to 8 GHz. The use of high-frequency maps is advantageous for identifying regions of synchrotron ageing. Interestingly, for two XRGs in that sample, namely 3C 223.1 and 3C 403, \citet{Rottmann01} reported an anomalous spectral index distribution, with the wings exhibiting a flatter radio spectrum compared to the primary lobes. He also found these two sources to have the smallest spectral ages in his XRG sample. \citet{Dennett-Thorpe02} confirmed a flatter spectrum for the wings in 3C 223.1, but found the spectral difference to be marginal ($\alpha_{\rm lobe}$ - $\alpha_{\rm wing}$ \simi 0.08). A similar `tendency' was reported by \citet{Mack05}, based on their spectral index map (74 - 1400 MHz), with a 45\arcsec ~ beam which could, however, scarcely resolve the wings. On the other hand, a distinctly flatter spectrum of the wings in 3C 223.1 was reported by \citet{Lal05}, in spite of their use of maps made at metre-wavelengths (240 - 610 MHz), where spectral steepening is expected to be less pronounced. These rather dissonant findings about the significance level of spectral flattening in the wings have provided us the impetus to take a fresh look into the reported spectral peculiarity of this $z=$  0.1075 XRG. 

We have taken advantage of the recently available LOFAR maps of 3C 223.1 at 144 MHz with 6\arcsec and 20\arcsec ~ beams (LoTSS-DR2\footnote{\url{https://lofar-surveys.org/dr2_release.html}} ; \citealt{LOTSSDR2}), in conjunction with the VLA images at C-band and X-band, obtained by us from the NRAO VLA Archive Survey \footnote{\url{http://www.vla.nrao.edu/astro/nvas/}}. These VLA D-array maps at C and X-bands have beamwidths of 14.2\arcsec $\times$ 11.7\arcsec, and 8.2\arcsec $\times$ 6.7\arcsec, respectively. 

\begin{figure*}[ht!]
\centering
\includegraphics[scale=0.11]{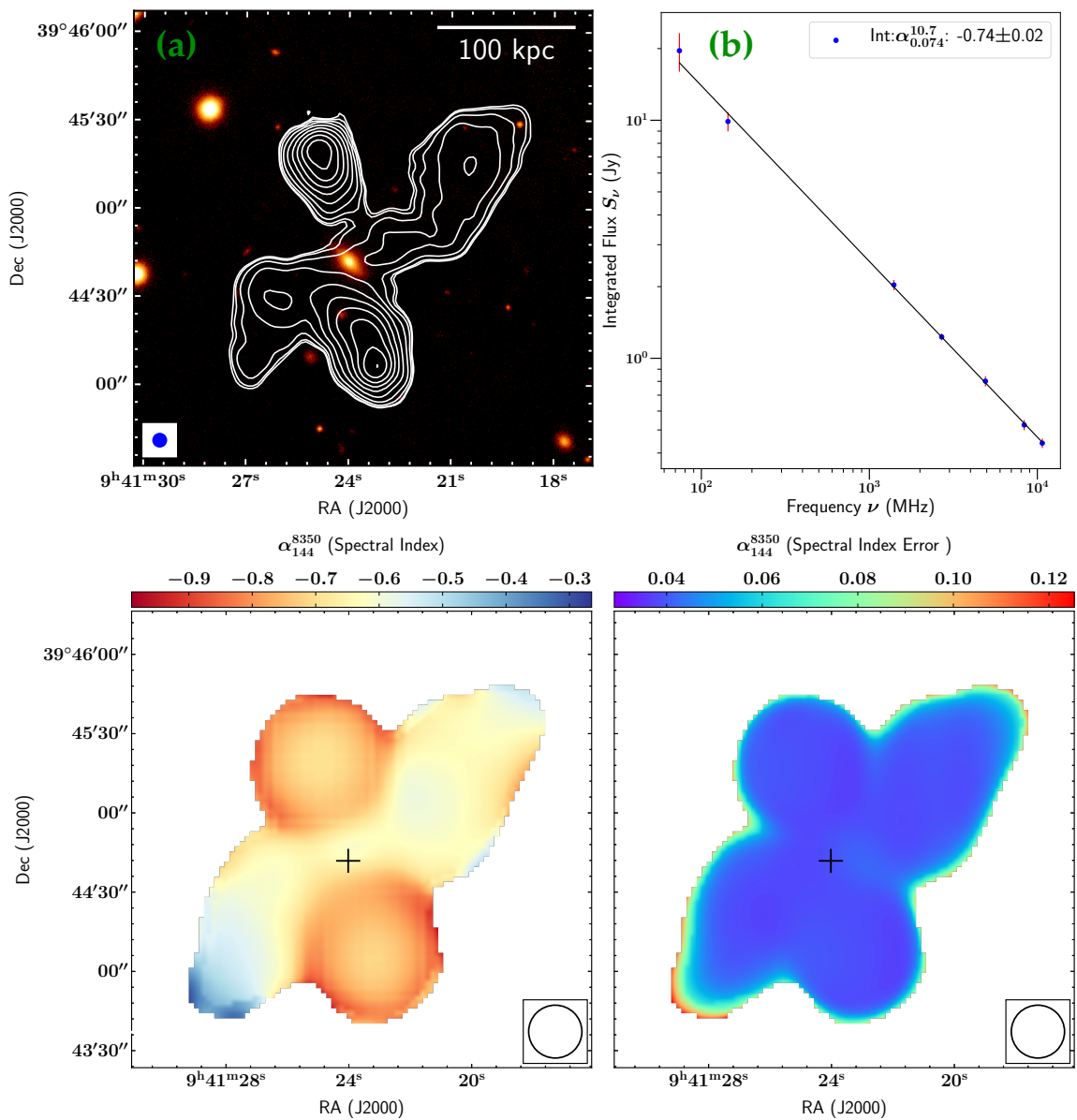}
\caption{{\tiny Radio imaging and spectral data for 3C223.1. (a) LoTSS DR2 144 MHz radio contours (levels: 0.0085,0.0097,0.0134,0.0213,0.0384,0.0754,0.1549,0.3263,0.6955, and 1.4909 Jy~beam$^{-1}$) of the 6\arcsec map overlaid on PanSTARRS r band image. (b): Integrated spectral index fit from 144 to 10700 MHz. (c): Three frequencies (144, 4910, and 8350 MHz) spectral index map  and error map (d) of 3C223.1. The beam of 20\arcsec $\times$ 20\arcsec ~ is shown at the bottom right corner. The host galaxy location is shown by the black marker.}}
\label{fig:lofarsifit}
\end{figure*}

\begin{figure}
\includegraphics[scale=0.18]{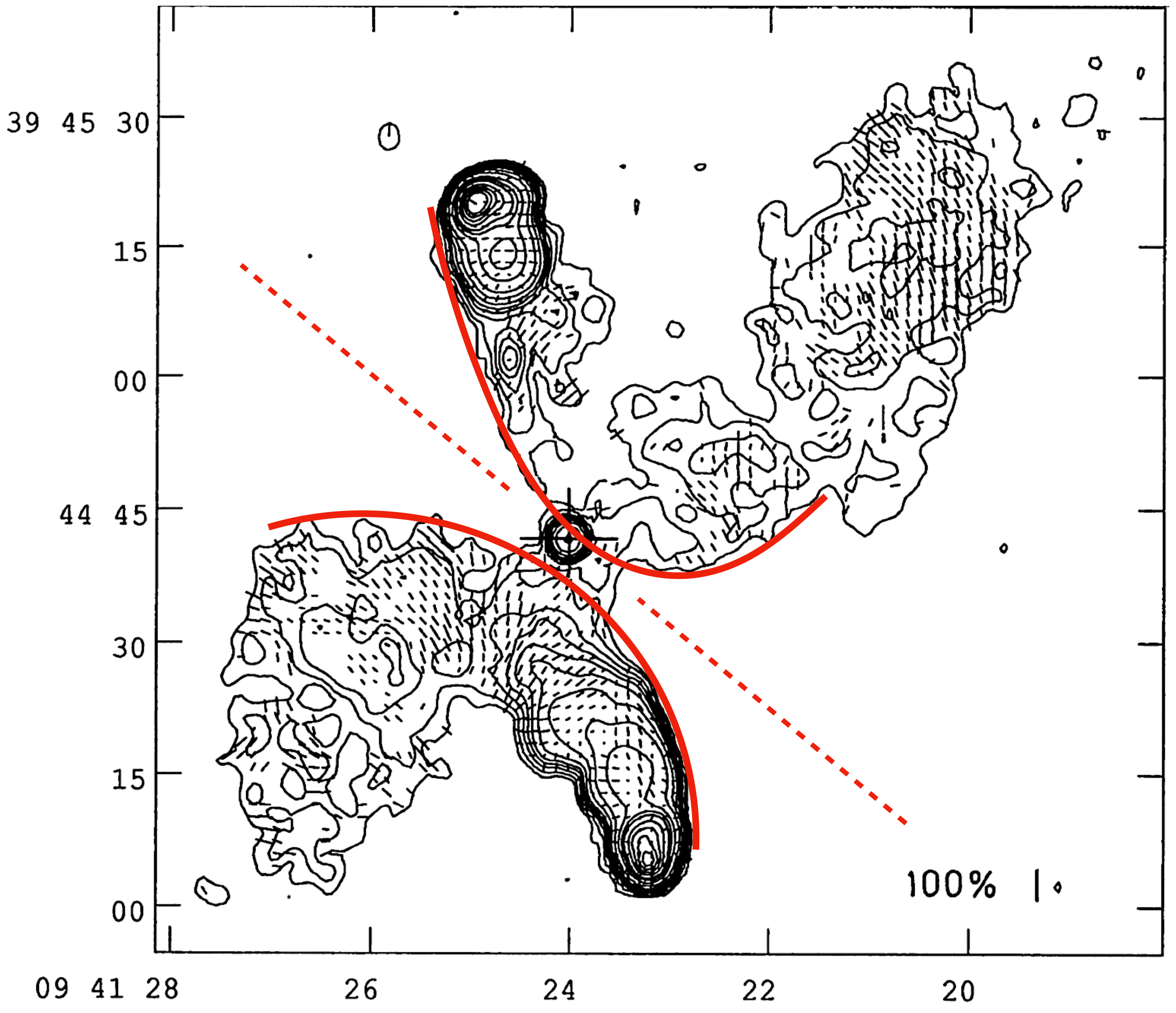}
\caption{{\tiny Double-boomerang morphology of XRG 223.1 can be clearly seen in the above figure based on Figure 4b of \citet{Black1992}, where VLA 8 GHz 2.5\arcsec ~ image contours of 3C223.1 are shown with polarisation vectors. The contours are drawn at (-4,4,6,9,12,15,20,30,50,100,150,200,300,400,500, and 600) $\times$ 55~$\muup$Jy~beam$^{-1}$ (rms or $\rm \sigma_{I}$). For the polarized intensity, the $\rm \sigma_{P}$ is \simi36~$\muup$Jy~beam$^{-1}$ and the vectors are drawn only at regions which have a surface-brightness $>$4$\sigma_{I}$. We have marked outline of the `double boomerang' in red colour along with the dashed line showing 40\deg ~ position angle of the disk of the host galaxy (Fig.\ \ref{fig:lofarsifit}a).}}
\label{fig:vlapol}
\end{figure} 

\vspace{-0.3cm}
\section{Spectral index mapping of the XRG 3C 223.1} \label{sec:2}
The above-mentioned LoTSS-DR2 map at 144 MHz was combined with the VLA maps at 4910 MHz and 8350 MHz, following the method given by \citet{duy17} for generating the three-frequencies spectral index map with 20\arcsec matched beam. These VLA D-array maps at 4.91 and 8.35 GHz essentially recover the entire radio structure of this XRG, since their integrated flux densities of 0.80$\pm$0.04 Jy and 0.52$\pm$0.03 Jy, respectively, are in full accordance with the integrated spectrum 
(see Fig.\ \ref{fig:lofarsifit}b and Table\ \ref{tab:radioimtab}). 

This is consistent with the fact that even at 8.35 GHz, dense UV coverage of the VLA D-array data used here extends down to 35 m, which is enough to pick up the entire structure of this bright 2.0\arcmin ~ radio source. Fig.\ \ref{fig:lofarsifit} (c-d) displays the derived spectral index and error maps, based on a combination of high sensitivity, resolution, and frequency range ( 1 : 58 ) that is unmatched for this XRG. Spectral flattening is distinctly visible towards each wing.
In the primary lobes, the northern and southern hot spots have  $\alpha_{144}^{8350}$ = -0.70$\pm$0.04 and -0.73 $\pm$0.04, respectively, and their associated (primary) lobes exhibit a spectral steepening by $\Delta\alpha$ \simi 0.1. However, going further into the wings, the spectral gradient reverses sign and the spectrum turns markedly flatter along the ridge line, right up to the wing's edge. Interestingly, spectrum in certain parts of the wings is even flatter than it is at the hot spots. This confirms, with a much higher level of precision and in much greater spatial detail, the original result from \cite{Rottmann01} and it is also consistent with the findings of \citet{Lal05}. This point is further discussed in the next section.

\begin{table}[htbp]
\setlength{\tabcolsep}{4pt}
\caption{{\tiny Integrated flux densities of 3C223.1 (Fig.\ \ref{fig:lofarsifit}b). Reference: 1)Present work, based on LoTSS-DR2 \citep{LOTSSDR2} \& NVSS \citep{nvss},  2)\citet{Kellermann68}, 3) \citet{Kellermann73}, 4) \citet{Black1992}, 5) \citet{Mack05}. }}\label{tab:radioimtab}
\begin{tabular}{lcccc}
\hline
 Frequency & Flux  & Telescope &Beam & Ref \\ 
  (MHz)    & (Jy)  &           &     &  \\ 

\hline
74 & 19.7$\pm$3.6  & VLA-A  & 24\arcsec  & 5\\
144 & 9.7$\pm$0.9 & LoFAR & 20\arcsec & 1\\
1400 & 2.0$\pm$0.1 &VLA-C-D  & 45\arcsec & 1\\
2695 & 1.23$\pm$0.04 & NRAO-140ft & 11.3\arcmin $\times$10.5\arcmin &  2\\
4910 & 0.80$\pm$0.04 & VLA-D & 20\arcsec &  4\\
8350 & 0.52$\pm$0.03 & VLA-D & 20\arcsec &  4\\
10700 & 0.31$\pm$0.03 & Effelsberg & 2.85\arcmin &  3\\

\hline
\end{tabular}
\end{table}

\vspace{-0.7cm}
\section{Discussion}\label{sec:Disc}
\begin{figure}
\centering
\includegraphics[scale=0.38]{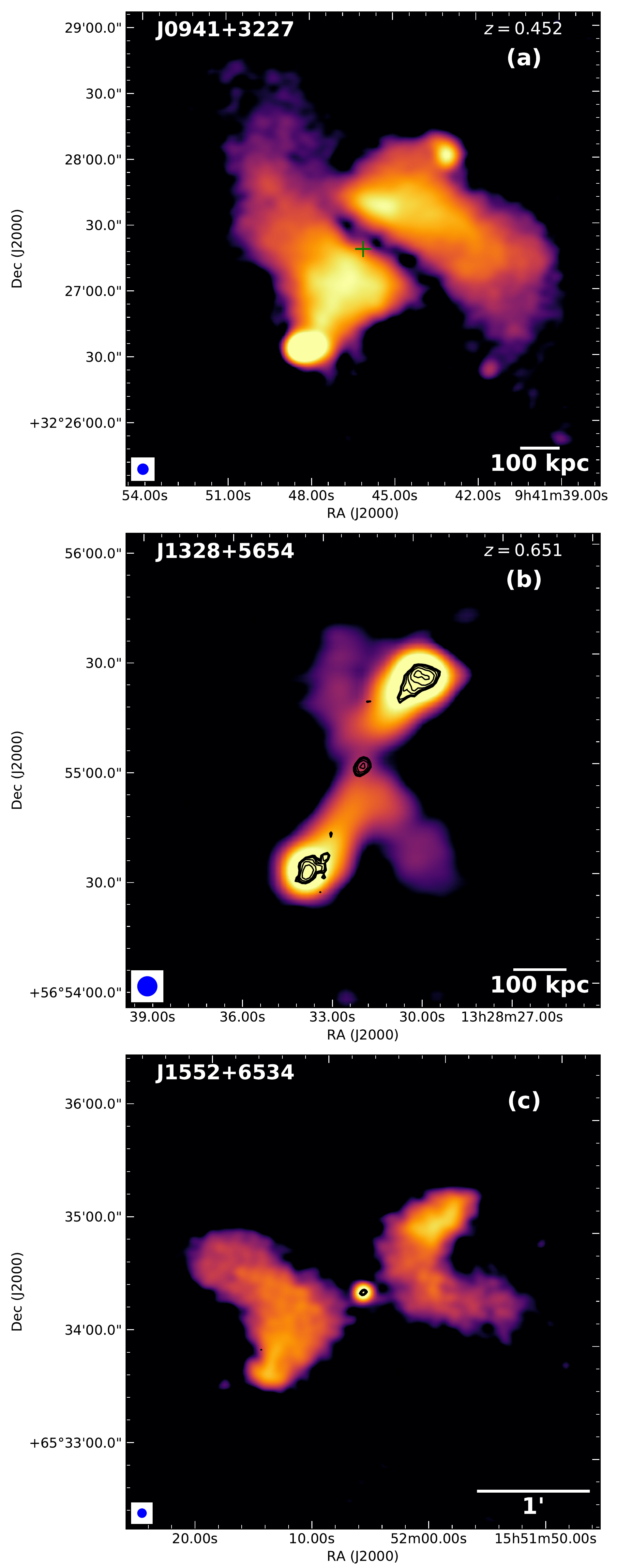}
\caption{{\tiny Three spectacular XRGs culled by us from LoTSS-DR2: 6\arcsec ~ LoTSS 144 MHz  images overlaid with VLASS 2.5\arcsec contours, where only the radio emission greater than 3$\sigma$ is shown. More details given in Table\ \ref{tab:3xrgs}. For J0941+3227, the radio core coincident (not shown here) with the host galaxy is quite faint and detected at 1$\sigma$ in both FIRST and VLASS, marked with `+' symbol in panel a.}}
\label{fig:3xrgs}
\end{figure}

The spectral index map presented in Fig.\ \ref{fig:lofarsifit}c clearly establishes XRG 3C 223.1 (J094124.028+394441.95) as the prime example of an XRG whose wings have a flatter radio spectrum than the primary lobes, thus challenging the currently popular models of XRGs, including the back-flow diversion model mentioned in Sect.\ \ref{sec:intro}. A potentially useful hint for the origin of this anomalous spectral gradient comes from the recent MeerKAT observations of the XRG PKS 2014-55, dubbed as `double-boomerang' XRG (dbXRG), which is hosted by a Seyfert 2 elliptical galaxy located in a poor group of galaxies at $z$ = 0.06063 (\citealt{cotton20} and references therein). Although this giant XRG is currently the leading exponent of the `double-boomerang' morphology, a few other XRGs with a similar appearance have been reported \citep[e.g.][]{Lal19}. Examples of this include the prototypical XRG, 3C 315 itself \citep{hogbom74}, as well as the newly recognised case of XRG J1552+6534, whose LoTSS-DR2 image is presented in Fig.\ \ref{fig:3xrgs}, resembling a `double crescent'. It may be noted that even for 3C 223.1, the existing 8.3 GHz VLA map with a 2.5\arcsec ~ beam \citep{Black1992} exhibits a turn-around of the back flow in each lobe, akin to the double-boomerang morphology (see, Fig.\ \ref{fig:vlapol}).  The map also shows that the magnetic field is aligned with the edges of the radio lobes (see, also, \citealt{Dennett-Thorpe02}). Probably due to the much greater relative spatial resolution available for the giant dbXRG PKS 2014-55, a sharp-edged faint radio cocoon of typical width \simi50 kpc has been detected around both its lobe pairs and the two radio cocoons appear to almost touch each other near their apexes where the elliptical host galaxy is situated  (Fig. 5 in \citealt{cotton20}). The cocoons appear to act as a sheath around the backflow both before and after its deflection. However, due to their faintness, the magnetic field geometry inside the radio cocoons is essentially unknown. Although the exceptionally high relative spatial resolution afforded by the giant size of that dbXRG is not yet achievable for other dbXRGs, it seems reasonable to expect that the backflows in them are also surrounded by similar protective radio cocoons.

In the case of 3C 223.1 as well, the two boomerangs are seen to approach each other to within \simi 2\arcsec (\simi4.3 kpc), near the location of the host elliptical galaxy which is known to possess a conspicuous dusty disk extending across the stellar body of the optical galaxy (Fig.\ \ref{fig:lofarsifit}a, Fig.\ \ref{fig:vlapol}), at a position angle of \simi40\deg. The dusty disk was detected in a \textit{Hubble Space Telescope} snapshot survey of radio galaxies \citep{deKoff96}. This disk of large extent and right orientation may well be playing a significant role in blocking and deflecting the hydrodynamic backflow streaming through the two primary lobes. The compression of the magnetic field of the disk (and, possibly, of its synchrotron halo) by the impact of the collimated backflow could be contributing to the powerful push needed to transform the obliquely incident backflow into a boomerang shape. Post-rebound, the backflow propagation is guided by the steepest pressure gradient in the ISM of the host galaxy, as envisioned in \citet{leahy84}. A similar magnetic rebound may have contributed to the formation of the classic double boomerang in the giant dbXRG PKS 2014-55 where the backflow has been posited to impinge upon an ellipsoidal gaseous interstellar medium (ISM) of the host galaxy, with a required extent of \simi150 kpc \citep{cotton20}. We note that even in this elliptical Seyfert-2 galaxy, a nearly edge-on dusty disk has been detected which too is oriented nearly along the symmetry plane of the double boomerang \citep{Abbott2018,cotton20}. However, in order to effectively contribute to the backflow rebounding observed in this giant XRG, the gaseous disk would have to be larger by one order of magnitude than its detected extent of about 12 kpc. Such a possibility has not been ruled out, however. Sensitive H{\sc i} imaging of non-cluster early-type galaxies have revealed cases in which an H{\sc i} disk extends over several tens of kpc, which was probably acquired from one or more approaching gas-rich galaxies, and such H{\sc i} disks are prone to having kpc-scale central disks of dusty molecular gas \citep{Serra2012,Yildiz2020}.   

Here, it is pertinent to recall an independent evidence for the role of magnetic tension, which has emerged from recent MeerKAT observations of the radio galaxy MRC 0600-399 in Abell 3376, followed-up with numerical simulations. Based on this information, \citet{Chibueze21} have argued that even the observed sharp bending of the relatively powerful jets of this wide-angle-tail (WAT) radio galaxy has taken place due to the jets encountering the tension of a compressed layer of external magnetic field. More generally, the backflowing synchrotron plasma of a lobe could also be diverted upon hitting a flattened gaseous structures, such as a sheet or filament of the cosmic web, as recently proposed for the case of the giant radio galaxy GRG 0503-286 (\citealt{Dabhade22}, see also, \citealt{GK09}). A spectacular example of a flattened gaseous obstruction between the two lobes can be seen in the recent LoTSS-DR2 image of a large XRG (J0941+3227 of size \simi 0.6 Mpc), whose lobes appear separated by a `linear' gap of average width \simi25 kpc (Fig.\ \ref{fig:3xrgs}a).

\begin{table*}[htbp]
\centering
\setlength{\tabcolsep}{4pt}

\caption{{\tiny  Coordinates of the host galaxies of 3 spectacular XRGs selected by us from LoTSS-DR2 (Fig.\ \ref{fig:3xrgs}). The redshift ($z$), taken from the Sloan Digital Sky Survey (SDSS; \citealt{sdssyork}), is spectroscopic for J0941+3227 and photometric for J1328+5654. The sizes refer to the separation between the two hotspots (in the primary lobes). S$_{\rm 144}$ is the integrated flux density at 144 MHz (LoTSS-DR2) and P$_{144}$ the corresponding total radio luminosity. $\sigma_{\rm map}$ is the rms noise in the maps (Fig.\ \ref{fig:3xrgs}). Throughout this paper, we have adopted a flat cosmology with parameters $\rm \Omega_m$ = 0.27, and a Hubble constant of H$_0$ = 71 km s$^{-1}$ Mpc$^{-1}$. }}\label{tab:3xrgs}

\begin{tabular}{lcccccccc}
\hline
 Source & R.A (J2000)  & Dec (J2000) & $z$ & Size     &   Size  & S$_{\rm 144}$ & P$_{144}$ &$\sigma_{\rm map}$\\ 
        &      &     &     &(\arcsec) &   (kpc) &     (Jy)       &  ($\times 10^{26}$ W~Hz$^{-1}$) &($\muup$Jy~beam$^{-1}$)\\ 

\hline

J0941+3227 & 09:41:46.10        & +32:27:18.64  & 0.45261$\pm$0.00005   & 110     & 634   & 0.90$\pm$0.09 & 6.0&103\\

J1328+5654 & 13:28:31.82        & +56:54:59.41  & 0.651$\pm$0.073                   & 61 & 427   & 0.34$\pm$0.03& 5.2 & 65\\
J1552+6534 & 15:52:06.34        & +65:34:24.57  & -                                 & 128        & -        & 0.25$\pm$0.03 &- & 91\\

\hline
\end{tabular}
\end{table*}

\vspace{-0.3cm}
\subsection{Clues to the flatter radio spectrum of the wings}
In this section, we summarise some observational clues bearing on the question of the flatter radio spectrum of XRG wings (compared to the primary lobes), for which 3C 223.1 is thus demonstrated to be a proto-type. In this dbXRG, the clear reversal of radio spectral gradient following the deflection of the backflow into the wings (Sect.\ \ref{sec:2}) stands in sharp contrast to the monotonous spectral steepening along the backflow, which is typical of edge-brightened double radio sources. We propose that the spectral flattening observed towards the wings in 3C 223.1 is linked to particle acceleration (or re-acceleration) during the process of rebounding of the backflow. Plausibly, this could occur as the backflow impinges upon the disk (and its likely synchrotron halo) and encounters the tension of their magnetic field lines compressed by the impact of the backflow. We may recall that in both aforementioned examples, dbXRG PKS 2014-55 \citep{cotton20} and the cluster radio source MRC 0600-399 \citep{Chibueze21}, localised regions of enhanced radio emission accompanied by spectral flattening have actually been observed near the areas where a powerful collimated flow of the synchrotron plasma (jet or backflow) undergoes a sharp bending or rebound. Such patches of flatter radio spectrum could either dilute or mask the effect of spectral steepening in the ageing synchrotron plasma deflected into the XRG wings, or, in the extreme case, may even cause spectral flattening in the wings. Recent MHD simulations, reported in \citep{Chibueze21}, have shown that when a jet flow encounters the tension of magnetic field lines in a compressed layer, an efficient conversion of the magnetic energy into relativistic particles via magnetic reconnection can occur; then, the relativistic particles, accelerated in situ, get transported along the deflected stream of synchrotron plasma \citep[see, also,][]{Giri2022}. A similar process may be occurring at a significant level in the XRGs whose wings do not show spectral steepening or even exhibit a spectral flattening; for instance, in the rare case of XRG 3C 223.1. In view of this, it would be desirable to extend the spectral index mapping to the several other XRGs which are candidates for wings having flatter radio spectra compared to the primary lobes, based on their metre wavelength imaging observations \citep{Lal19}. It is important to extend their spectral mapping to centimetre wavelengths, where spectral steepening due to synchrotron losses should be more pronounced. Also, it would be instructive to look for signs of spectral flattening in regions where the jet flow appears to undergo a deflection upon colliding with an obstruction such as a galaxy shell. This possibility seems particularly relevant to the XRGs in which the wings take off sharply from the primary lobes at large distances ( $>>$ 10 kpc) from the parent galaxy, which is well beyond the likely interstellar medium of the parent galaxy \citep{xrggk12,Joshi19}. Observational evidence for jet-shell interactions \citep{GKCHITRE83} has been reported for the nearest radio galaxy Centaurus A \citep{GK84,GK10}. 

\vspace{-0.3cm}
\subsection{A new kind of lobe symmetry in XRGs}
Conceptually, the most straightforward explanation for XRGs -- perhaps inspired by the XRGs whose wings do not exhibit a steeper radio spectrum compared to the primary lobes -- is that the central engine consists of a close binary of SMBH (see, \citealt{Begelman1980}),  each of which produces its own pair of lobes \citep{Lal05,Lal19}. However, several observations have questioned the general applicability of this model, including the ubiquitous absence (i) of parsec-scale nuclear radio jets pointing towards the wings and (ii) of terminal hot spots in the wings (see, \citealt{xrggk12} for a review of the models of XRGs). Another challenge to this hypothesis stems from the detection of a lateral offset between the ridge-lines of the two wings in some well-mapped XRGs \citep{GK2003}. We note that such an offset is problematic even for the basic spin-flip model (Sec.\ \ref{sec:intro}), however, that model can be reconciled in case the wings arise due to bending of the twin jets by the ISM of the host galaxy, which has been set in rotation during the orbital infall of a merging galaxy \citep{GK2003}. 

Here, we would like to draw attention to a new kind of morphological peculiarity, namely, a lateral offset between the pair of 'primary' radio lobes that extend parallel to each other. The case of XRG (J1328+5654) exemplifying such an anomalous morphology is shown in Fig.\ \ref{fig:3xrgs}b, featuring the LoFAR image and contours from the 3 GHz Very Large Array Sky Survey (VLASS; \citealt{vlass}). This specific morphology of the primary lobes is puzzling, since they are widely believed to be directly fed by the bipolar jets emanating from the nucleus, unlike the wings (see, e.g. \citealt{cotton20} and references therein). Unfortunately, the existing radio maps of this XRG lack the spatial resolution and sensitivity to trace the detailed trajectories of its jets, from the nucleus to the terminal hot spots seen in the primary lobes. It would be instructive to obtain this vital information through sensitive, high-resolution radio imaging in order to unravel how the bipolar jets in such XRGs undergo bending and how the backflows in the two lobes remain parallel to each other while being laterally offset.

\section{Conclusions} \label{sec:conc}

Taking advantage of the recent availability of the LoTSS-DR2 map of the X-shaped radio galaxy (XRG) 3C 223.1 at 144 MHz and using its archival VLA observations at 4.9 GHz and 8.4 GHz, we mapped the radio spectral index distribution across this XRG with unprecedented precision and spatial detail. This firmly establishes it as a prime example of an XRG in which the radio wings exhibit a distinctly flatter spectrum than the primary lobes, setting aside the debate over the level of this spectral anomaly. Evidence is also presented in support of this XRG having a `double boomerang' type radio morphology. Based on existing observational clues, we suggest that the flatter spectrum of the wings in this XRG manifests an extreme case of in situ acceleration and energisation of relativistic particles as the collimated hydrodynamical backflow of synchrotron plasma in the primary lobes impinges obliquely upon the prominent gaseous disk (and its likely synchrotron plasma halo) within the host galaxy and rebounds due to the tension of its magnetic field lines, which are compressed by the impact of the collimated backflows from opposite sides. Lastly, we have drawn attention to a new and intriguing morphological symmetry whereby the two primary lobes of an XRG, although parallel to each other, have a distinct lateral offset. Explaining this morphological anomaly appears more challenging than for similar morphological pattern found for the wings in several XRGs.

\section*{Acknowledgements}
GK acknowledges a Senior Scientist fellowship of the Indian National Science Academy. We would like to dedicate this work to late Prof. S.M. Chitre who, together with one of the authors (GK) introduced the concept of jet-shell interactions in radio galaxies \citep{GKCHITRE83}. The acknowledgment for LoTSS data usage can be found at \url{https://lofar-surveys.org/credits.html}.
The VLA archival data was obtained via the NRAO VLA Archive Survey, (c) AUI/NRAO.
We acknowledge that this work has made use of \textsc{aplpy} \citep{apl}. 

\bibliographystyle{aa} 
\bibliography{XRG_3C223.bib}

\begin{thebibliography}{39}
\expandafter\ifx\csname natexlab\endcsname\relax\def\natexlab#1{#1}\fi

\bibitem[{{Abbott} {et~al.}(2018){Abbott}, {Abdalla}, {Allam}, {Amara},
  {Annis}, {Asorey}, {Avila}, {Ballester}, {Banerji}, {Barkhouse}, {Baruah},
  {Baumer}, {Bechtol}, {Becker}, {Benoit-L{\'e}vy}, {Bernstein}, {Bertin},
  {Blazek}, {Bocquet}, {Brooks}, {Brout}, {Buckley-Geer}, {Burke}, {Busti},
  {Campisano}, {Cardiel-Sas}, {Carnero Rosell}, {Carrasco Kind}, {Carretero},
  {Castander}, {Cawthon}, {Chang}, {Chen}, {Conselice}, {Costa}, {Crocce},
  {Cunha}, {D'Andrea}, {da Costa}, {Das}, {Daues}, {Davis}, {Davis}, {De
  Vicente}, {DePoy}, {DeRose}, {Desai}, {Diehl}, {Dietrich}, {Dodelson},
  {Doel}, {Drlica-Wagner}, {Eifler}, {Elliott}, {Evrard}, {Farahi}, {Fausti
  Neto}, {Fernandez}, {Finley}, {Flaugher}, {Foley}, {Fosalba}, {Friedel},
  {Frieman}, {Garc{\'\i}a-Bellido}, {Gaztanaga}, {Gerdes}, {Giannantonio},
  {Gill}, {Glazebrook}, {Goldstein}, {Gower}, {Gruen}, {Gruendl}, {Gschwend},
  {Gupta}, {Gutierrez}, {Hamilton}, {Hartley}, {Hinton}, {Hislop}, {Hollowood},
  {Honscheid}, {Hoyle}, {Huterer}, {Jain}, {James}, {Jeltema}, {Johnson},
  {Johnson}, {Kacprzak}, {Kent}, {Khullar}, {Klein}, {Kovacs}, {Koziol},
  {Krause}, {Kremin}, {Kron}, {Kuehn}, {Kuhlmann}, {Kuropatkin}, {Lahav},
  {Lasker}, {Li}, {Li}, {Liddle}, {Lima}, {Lin}, {L{\'o}pez-Reyes}, {MacCrann},
  {Maia}, {Maloney}, {Manera}, {March}, {Marriner}, {Marshall}, {Martini},
  {McClintock}, {McKay}, {McMahon}, {Melchior}, {Menanteau}, {Miller},
  {Miquel}, {Mohr}, {Morganson}, {Mould}, {Neilsen}, {Nichol}, {Nogueira},
  {Nord}, {Nugent}, {Nunes}, {Ogando}, {Old}, {Pace}, {Palmese},
  {Paz-Chinch{\'o}n}, {Peiris}, {Percival}, {Petravick}, {Plazas}, {Poh},
  {Pond}, {Porredon}, {Pujol}, {Refregier}, {Reil}, {Ricker}, {Rollins},
  {Romer}, {Roodman}, {Rooney}, {Ross}, {Rykoff}, {Sako}, {Sanchez}, {Sanchez},
  {Santiago}, {Saro}, {Scarpine}, {Scolnic}, {Serrano}, {Sevilla-Noarbe},
  {Sheldon}, {Shipp}, {Silveira}, {Smith}, {Smith}, {Smith}, {Soares-Santos},
  {Sobreira}, {Song}, {Stebbins}, {Suchyta}, {Sullivan}, {Swanson}, {Tarle},
  {Thaler}, {Thomas}, {Thomas}, {Troxel}, {Tucker}, {Vikram}, {Vivas},
  {Walker}, {Wechsler}, {Weller}, {Wester}, {Wolf}, {Wu}, {Yanny}, {Zenteno},
  {Zhang}, {Zuntz}, {DES Collaboration}, {Juneau}, {Fitzpatrick}, {Nikutta},
  {Nidever}, {Olsen}, {Scott}, \& {NOAO Data Lab}}]{Abbott2018}
{Abbott}, T.~M.~C., {Abdalla}, F.~B., {Allam}, S., {et~al.} 2018, \apjs, 239,
  18

\bibitem[{{Alexander} \& {Leahy}(1987)}]{Alexander87}
{Alexander}, P. \& {Leahy}, J.~P. 1987, \mnras, 225, 1

\bibitem[{{Begelman} {et~al.}(1980){Begelman}, {Blandford}, \&
  {Rees}}]{Begelman1980}
{Begelman}, M.~C., {Blandford}, R.~D., \& {Rees}, M.~J. 1980, \nat, 287, 307

\bibitem[{{Black} {et~al.}(1992){Black}, {Baum}, {Leahy}, {Perley}, {Riley}, \&
  {Scheuer}}]{Black1992}
{Black}, A.~R.~S., {Baum}, S.~A., {Leahy}, J.~P., {et~al.} 1992, \mnras, 256,
  186

\bibitem[{{Capetti} {et~al.}(2002){Capetti}, {Zamfir}, {Rossi}, {Bodo},
  {Zanni}, \& {Massaglia}}]{Capetti02}
{Capetti}, A., {Zamfir}, S., {Rossi}, P., {et~al.} 2002, \aap, 394, 39

\bibitem[{{Chibueze} {et~al.}(2021){Chibueze}, {Sakemi}, {Ohmura}, {Machida},
  {Akamatsu}, {Akahori}, {Nakanishi}, {Parekh}, {van Rooyen}, \&
  {Takeuchi}}]{Chibueze21}
{Chibueze}, J.~O., {Sakemi}, H., {Ohmura}, T., {et~al.} 2021, \nat, 593, 47

\bibitem[{{Condon} {et~al.}(1998){Condon}, {Cotton}, {Greisen}, {Yin},
  {Perley}, {Taylor}, \& {Broderick}}]{nvss}
{Condon}, J.~J., {Cotton}, W.~D., {Greisen}, E.~W., {et~al.} 1998, \aj, 115,
  1693

\bibitem[{{Cotton} {et~al.}(2020){Cotton}, {Thorat}, {Condon}, {Frank},
  {J{\'o}zsa}, {White}, {Deane}, {Oozeer}, {Atemkeng}, {Bester}, {Fanaroff},
  {Kupa}, {Smirnov}, {Mauch}, {Krishnan}, \& {Camilo}}]{cotton20}
{Cotton}, W.~D., {Thorat}, K., {Condon}, J.~J., {et~al.} 2020, \mnras, 495,
  1271

\bibitem[{{Dabhade} \& {Gopal-Krishna}(2022)}]{Dabhade22}
{Dabhade}, P. \& {Gopal-Krishna}. 2022, \aap, 660, L10

\bibitem[{{de Koff} {et~al.}(1996){de Koff}, {Baum}, {Sparks}, {Biretta},
  {Golombek}, {Macchetto}, {McCarthy}, \& {Miley}}]{deKoff96}
{de Koff}, S., {Baum}, S.~A., {Sparks}, W.~B., {et~al.} 1996, \apjs, 107, 621

\bibitem[{{Dennett-Thorpe} {et~al.}(2002){Dennett-Thorpe}, {Scheuer}, {Laing},
  {Bridle}, {Pooley}, \& {Reich}}]{Dennett-Thorpe02}
{Dennett-Thorpe}, J., {Scheuer}, P.~A.~G., {Laing}, R.~A., {et~al.} 2002,
  \mnras, 330, 609

\bibitem[{{Giri} {et~al.}(2022){Giri}, {Vaidya}, {Rossi}, {Bodo}, {Mukherjee},
  \& {Mignone}}]{Giri2022}
{Giri}, G., {Vaidya}, B., {Rossi}, P., {et~al.} 2022, \aap, 662, A5

\bibitem[{{Gopal-Krishna} {et~al.}(2012){Gopal-Krishna}, {Biermann}, {Gergely},
  \& {Wiita}}]{xrggk12}
{Gopal-Krishna}, {Biermann}, P.~L., {Gergely}, L.~{\'A}., \& {Wiita}, P.~J.
  2012, Research in Astronomy and Astrophysics, 12, 127

\bibitem[{{Gopal-Krishna} {et~al.}(2003){Gopal-Krishna}, {Biermann}, \&
  {Wiita}}]{GK2003}
{Gopal-Krishna}, {Biermann}, P.~L., \& {Wiita}, P.~J. 2003, \apjl, 594, L103

\bibitem[{{Gopal-Krishna} \& {Chitre}(1983)}]{GKCHITRE83}
{Gopal-Krishna} \& {Chitre}, S.~M. 1983, \nat, 303, 217

\bibitem[{{Gopal-Krishna} \& {Saripalli}(1984)}]{GK84}
{Gopal-Krishna} \& {Saripalli}, L. 1984, \aap, 141, 61

\bibitem[{{Gopal-Krishna} \& {Wiita}(2009)}]{GK09}
{Gopal-Krishna} \& {Wiita}, P.~J. 2009, \na, 14, 51

\bibitem[{{Gopal-Krishna} \& {Wiita}(2010)}]{GK10}
{Gopal-Krishna} \& {Wiita}, P.~J. 2010, \na, 15, 96

\bibitem[{{Hoang} {et~al.}(2017){Hoang}, {Shimwell}, {Stroe}, {Akamatsu},
  {Brunetti}, {Donnert}, {Intema}, {Mulcahy}, {R{\"o}ttgering}, {van Weeren},
  {Bonafede}, {Br{\"u}ggen}, {Cassano}, {Chy{\.z}y}, {En{\ss}lin}, {Ferrari},
  {de Gasperin}, {Gu}, {Hoeft}, {Miley}, {Orr{\'u}}, {Pizzo}, \&
  {White}}]{duy17}
{Hoang}, D.~N., {Shimwell}, T.~W., {Stroe}, A., {et~al.} 2017, \mnras, 471,
  1107

\bibitem[{{Hodges-Kluck} {et~al.}(2010){Hodges-Kluck}, {Reynolds}, {Cheung}, \&
  {Miller}}]{Hodges-Kluck10}
{Hodges-Kluck}, E.~J., {Reynolds}, C.~S., {Cheung}, C.~C., \& {Miller}, M.~C.
  2010, \apj, 710, 1205

\bibitem[{{H\"{o}gbom}(1979)}]{Hogbom79}
{H\"{o}gbom}, J.~A. 1979, \aaps, 36, 173

\bibitem[{{H\"{o}gbom} \& {Carlsson}(1974)}]{hogbom74}
{H\"{o}gbom}, J.~A. \& {Carlsson}, I. 1974, \aap, 34, 341

\bibitem[{{Joshi} {et~al.}(2019){Joshi}, {Krishna}, {Yang}, {Shi}, {Yu},
  {Wiita}, {Ho}, {Wu}, {An}, {Wang}, {Subramanian}, \& {Yesuf}}]{Joshi19}
{Joshi}, R., {Krishna}, G., {Yang}, X., {et~al.} 2019, \apj, 887, 266

\bibitem[{{Kellermann} \& {Pauliny-Toth}(1973)}]{Kellermann73}
{Kellermann}, K.~I. \& {Pauliny-Toth}, I.~I.~K. 1973, \aj, 78, 828

\bibitem[{{Kellermann} {et~al.}(1968){Kellermann}, {Pauliny-Toth}, \&
  {Tyler}}]{Kellermann68}
{Kellermann}, K.~I., {Pauliny-Toth}, I.~I.~K., \& {Tyler}, W.~C. 1968, \aj, 73,
  298

\bibitem[{{Lacy} {et~al.}(2020){Lacy}, {Baum}, {Chandler}, {Chatterjee},
  {Clarke}, {Deustua}, {English}, {Farnes}, {Gaensler}, {Gugliucci},
  {Hallinan}, {Kent}, {Kimball}, {Law}, {Lazio}, {Marvil}, {Mao}, {Medlin},
  {Mooley}, {Murphy}, {Myers}, {Osten}, {Richards}, {Rosolowsky}, {Rudnick},
  {Schinzel}, {Sivakoff}, {Sjouwerman}, {Taylor}, {White}, {Wrobel},
  {Andernach}, {Beasley}, {Berger}, {Bhatnager}, {Birkinshaw}, {Bower},
  {Brandt}, {Brown}, {Burke-Spolaor}, {Butler}, {Comerford}, {Demorest}, {Fu},
  {Giacintucci}, {Golap}, {G{\"u}th}, {Hales}, {Hiriart}, {Hodge}, {Horesh},
  {Ivezi{\'c}}, {Jarvis}, {Kamble}, {Kassim}, {Liu}, {Loinard}, {Lyons},
  {Masters}, {Mezcua}, {Moellenbrock}, {Mroczkowski}, {Nyland}, {O'Dea},
  {O'Sullivan}, {Peters}, {Radford}, {Rao}, {Robnett}, {Salcido}, {Shen},
  {Sobotka}, {Witz}, {Vaccari}, {van Weeren}, {Vargas}, {Williams}, \&
  {Yoon}}]{vlass}
{Lacy}, M., {Baum}, S.~A., {Chandler}, C.~J., {et~al.} 2020, \pasp, 132, 035001

\bibitem[{{Lal} \& {Rao}(2005)}]{Lal05}
{Lal}, D.~V. \& {Rao}, A.~P. 2005, \mnras, 356, 232

\bibitem[{{Lal} {et~al.}(2019){Lal}, {Sebastian}, {Cheung}, \& {Pramesh
  Rao}}]{Lal19}
{Lal}, D.~V., {Sebastian}, B., {Cheung}, C.~C., \& {Pramesh Rao}, A. 2019, \aj,
  157, 195

\bibitem[{{Leahy} \& {Williams}(1984)}]{leahy84}
{Leahy}, J.~P. \& {Williams}, A.~G. 1984, \mnras, 210, 929

\bibitem[{{Mack} {et~al.}(2005){Mack}, {Vigotti}, {Gregorini}, {Klein},
  {Tschager}, {Schilizzi}, \& {Snellen}}]{Mack05}
{Mack}, K.~H., {Vigotti}, M., {Gregorini}, L., {et~al.} 2005, \aap, 435, 863

\bibitem[{{Merritt} \& {Ekers}(2002)}]{Merritt02}
{Merritt}, D. \& {Ekers}, R.~D. 2002, Science, 297, 1310

\bibitem[{{Robitaille} \& {Bressert}(2012)}]{apl}
{Robitaille}, T. \& {Bressert}, E. 2012, {APLpy: Astronomical Plotting Library
  in Python}

\bibitem[{{Rottmann}(2001)}]{Rottmann01}
{Rottmann}, H. 2001, PhD thesis, -

\bibitem[{{Serra} {et~al.}(2012){Serra}, {Oosterloo}, {Morganti}, {Alatalo},
  {Blitz}, {Bois}, {Bournaud}, {Bureau}, {Cappellari}, {Crocker}, {Davies},
  {Davis}, {de Zeeuw}, {Duc}, {Emsellem}, {Khochfar}, {Krajnovi{\'c}},
  {Kuntschner}, {Lablanche}, {McDermid}, {Naab}, {Sarzi}, {Scott}, {Trager},
  {Weijmans}, \& {Young}}]{Serra2012}
{Serra}, P., {Oosterloo}, T., {Morganti}, R., {et~al.} 2012, \mnras, 422, 1835

\bibitem[{{Shimwell} {et~al.}(2022){Shimwell}, {Hardcastle}, {Tasse}, {Best},
  {R{\"o}ttgering}, {Williams}, {Botteon}, {Drabent}, {Mechev}, {Shulevski},
  {van Weeren}, {Bester}, {Br{\"u}ggen}, {Brunetti}, {Callingham}, {Chy{\.z}y},
  {Conway}, {Dijkema}, {Duncan}, {de Gasperin}, {Hale}, {Haverkorn}, {Hugo},
  {Jackson}, {Mevius}, {Miley}, {Morabito}, {Morganti}, {Offringa}, {Oonk},
  {Rafferty}, {Sabater}, {Smith}, {Schwarz}, {Smirnov}, {O'Sullivan},
  {Vedantham}, {White}, {Albert}, {Alegre}, {Asabere}, {Bacon}, {Bonafede},
  {Bonnassieux}, {Brienza}, {Bilicki}, {Bonato}, {Calistro Rivera}, {Cassano},
  {Cochrane}, {Croston}, {Cuciti}, {Dallacasa}, {Danezi}, {Dettmar}, {Di
  Gennaro}, {Edler}, {En{\ss}lin}, {Emig}, {Franzen}, {Garc{\'\i}a-Vergara},
  {Grange}, {G{\"u}rkan}, {Hajduk}, {Heald}, {Heesen}, {Hoang}, {Hoeft},
  {Horellou}, {Iacobelli}, {Jamrozy}, {Jeli{\'c}}, {Kondapally}, {Kukreti},
  {Kunert-Bajraszewska}, {Magliocchetti}, {Mahatma}, {Ma{\l}ek}, {Mandal},
  {Massaro}, {Meyer-Zhao}, {Mingo}, {Mostert}, {Nair}, {Nakoneczny},
  {Nikiel-Wroczy{\'n}ski}, {Orr{\'u}}, {Pajdosz-{\'S}mierciak}, {Pasini},
  {Prandoni}, {van Piggelen}, {Rajpurohit}, {Retana-Montenegro}, {Riseley},
  {Rowlinson}, {Saxena}, {Schrijvers}, {Sweijen}, {Siewert}, {Timmerman},
  {Vaccari}, {Vink}, {West}, {Wo{\l}owska}, {Zhang}, \& {Zheng}}]{LOTSSDR2}
{Shimwell}, T.~W., {Hardcastle}, M.~J., {Tasse}, C., {et~al.} 2022, \aap, 659,
  A1

\bibitem[{{Worrall} {et~al.}(1995){Worrall}, {Birkinshaw}, \&
  {Cameron}}]{Worrall95}
{Worrall}, D.~M., {Birkinshaw}, M., \& {Cameron}, R.~A. 1995, \apj, 449, 93

\bibitem[{{Y{\i}ld{\i}z} {et~al.}(2020){Y{\i}ld{\i}z}, {Peletier}, {Duc}, \&
  {Serra}}]{Yildiz2020}
{Y{\i}ld{\i}z}, M.~K., {Peletier}, R.~F., {Duc}, P.~A., \& {Serra}, P. 2020,
  \aap, 636, A8

\bibitem[{{York} {et~al.}(2000){York}, {Adelman}, {Anderson}, {Anderson},
  {Annis}, {Bahcall}, {Bakken}, {Barkhouser}, {Bastian}, {Berman}, {Boroski},
  {Bracker}, {Briegel}, {Briggs}, {Brinkmann}, {Brunner}, {Burles}, {Carey},
  {Carr}, {Castander}, {Chen}, {Colestock}, {Connolly}, {Crocker}, {Csabai},
  {Czarapata}, {Davis}, {Doi}, {Dombeck}, {Eisenstein}, {Ellman}, {Elms},
  {Evans}, {Fan}, {Federwitz}, {Fiscelli}, {Friedman}, {Frieman}, {Fukugita},
  {Gillespie}, {Gunn}, {Gurbani}, {de Haas}, {Haldeman}, {Harris}, {Hayes},
  {Heckman}, {Hennessy}, {Hindsley}, {Holm}, {Holmgren}, {Huang}, {Hull},
  {Husby}, {Ichikawa}, {Ichikawa}, {Ivezi{\'c}}, {Kent}, {Kim}, {Kinney},
  {Klaene}, {Kleinman}, {Kleinman}, {Knapp}, {Korienek}, {Kron}, {Kunszt},
  {Lamb}, {Lee}, {Leger}, {Limmongkol}, {Lindenmeyer}, {Long}, {Loomis},
  {Loveday}, {Lucinio}, {Lupton}, {MacKinnon}, {Mannery}, {Mantsch}, {Margon},
  {McGehee}, {McKay}, {Meiksin}, {Merelli}, {Monet}, {Munn}, {Narayanan},
  {Nash}, {Neilsen}, {Neswold}, {Newberg}, {Nichol}, {Nicinski}, {Nonino},
  {Okada}, {Okamura}, {Ostriker}, {Owen}, {Pauls}, {Peoples}, {Peterson},
  {Petravick}, {Pier}, {Pope}, {Pordes}, {Prosapio}, {Rechenmacher}, {Quinn},
  {Richards}, {Richmond}, {Rivetta}, {Rockosi}, {Ruthmansdorfer}, {Sandford},
  {Schlegel}, {Schneider}, {Sekiguchi}, {Sergey}, {Shimasaku}, {Siegmund},
  {Smee}, {Smith}, {Snedden}, {Stone}, {Stoughton}, {Strauss}, {Stubbs},
  {SubbaRao}, {Szalay}, {Szapudi}, {Szokoly}, {Thakar}, {Tremonti}, {Tucker},
  {Uomoto}, {Vanden Berk}, {Vogeley}, {Waddell}, {Wang}, {Watanabe},
  {Weinberg}, {Yanny}, {Yasuda}, \& {SDSS Collaboration}}]{sdssyork}
{York}, D.~G., {Adelman}, J., {Anderson}, Jr., J.~E., {et~al.} 2000, \aj, 120,
  1579

\bibitem[{{Zier} \& {Biermann}(2001)}]{Zier01}
{Zier}, C. \& {Biermann}, P.~L. 2001, \aap, 377, 23

\end{thebibliography}


\end{document}